\documentstyle[11pt,newpasp,twoside,epsf]{article} 
\def\edcomment#1{\iffalse\marginpar{\raggedright\sl#1\/}\else\relax\fi} 
\reversemarginpar

\def\ref{\par\noindent\hangindent 20 pt}

\def\oc{$\omega$ Cen}
\def\Msun{M$_{\odot}$ }
\newcommand{\simgt}{\raisebox{-3.8pt}{$\;\stackrel{\textstyle > }{\sim }\;$}}

\begin{document}

\title{A Test of Newton's Law of Gravity in the Weak Acceleration Regime}

\author{Riccardo Scarpa, Gianni Marconi, Roberto Gilmozzi} 
\affil{European Southern Observatory, 3107 Alonso de Cordova, 
Santiago, Chile} 

\section{Introduction} 
Newton's law of gravity is routinely used to describe galaxies, even
though its validity has been fully verified only within the solar
system, in regimes of acceleration orders of magnitude stronger than
the ones typical of galaxies. Though we have plenty of reasons for
trusting Newton's law also in these weak regimes, 
there are strong observational evidence that all
spacecrafts in the periphery of the solar system are experiencing an
anomalous, unexplained acceleration toward the sun (Anderson et
al. 1998). Moreover, 
the modified Newtonian dynamics (MOND; Milgrom 1983,
Sanders \& McGaugh 2002), which  posit a
breakdown of Newton's law of gravity below few times $a_0 \sim
10^{-8}$ cm s$^{-2}$,
succeeds in explaining many properties of galaxies
and other astrophysical phenomena 
without invoking non-baryonic dark matter (DM).

Because of these empirical evidence, we decided to perform an experiment 
to test Newton's law of gravity. We focused on globular clusters
(GC) because they are the smallest virialized structure believed to be
DM free. This ensures GC's internal
dynamics should follow precisely the prediction of newton's law for
any acceleration, in particular below $a_0$. In the case a discrepancy would
be found, then DM can not be invoked to explain it, and 
Newton's law would be falsified.

\section{Results for a Pilot Experiment on $\omega$ Centauri}

We studied the outskirt of the GC \oc, which was selected for having known
internal proper motion (van Leeuwen et al. 2000) and because 
visible from Paranal.  To
reach gravitational acceleration below $10^{-7}$ cm s$^{-2}$, we have
selected 91 stars from van Leeuwen et al. (2000) at distance $>30$ pc
from the center and membership probability $>90$\%.  Radial
velocities with average accuracy of 0.8 km s$^{-1}$ were subsequently
obtained at the ESO Very Large Telescope (VLT) with UVES. Of the
selected candidates, 75 were found to be cluster members.

Combining our data with data from literature we 
trace the velocity dispersion $\sigma$ profile up to 45 pc from 
the center, finding $\sigma$ remains
large and basically constant at large radii (Fig. 1). 
Has evident from Fig. 1, the cluster is isotropic
so the use of radial velocity only to derive $\sigma$ 
does not limit the generality of our result. 
The profile flattens out for $r>25$ pc,
equivalent to an acceleration of gravity of $\sim 10^{-7}$ cm s$^{-2}$
(for cluster mass $4.2\times 10^6$ \Msun).  
This is comparable to the acceleration regime for
which dark matter starts to be relevant in galaxies.

The large $\sigma$ can be the result of tidal heating,
or can be due to a large number of binary stars in our sample. 
Alternatively, if \oc\ were a galaxy (Hilker \& Richtler 2000),
our result may indicate a substantial amount of DM survived 
all the tidal stripping that transformed the galaxy
into the cluster we see today. Though none of this possibility can be
ruled out by present data, all requires fine tuning of the relevant
parameter to explain the flattening of the velocity dispersion
profile. It is striking that \oc\ is hundreds of times smaller than 
a galaxy and still its
dispersion profile mimics precisely the one observed in
elliptical galaxies and explained invoking DM (Carollo et al. 1995).

Interestingly, for the GC Pal 13 a mass-to-light ratio
$M/L \simgt 11$ has been reported (Meylan 2001). Such high $M/L$, unique
among GCs, can be explained if Pal 13 is out of dynamical
equilibrium.  In view of
the result just found for \oc, we suggest the large 
$M/L$ is another manifestation  of a breakdown of Newton's law. 
For $M/L=3$ the acceleration of gravity 
is below $10^{-7}$ cm/s$^2$ all the way to the cluster center.  
It is therefore not surprisingly that Pal 13 appears dominated by DM. 

As a whole, we believe our result for \oc, the one for Pal 13, and the
anomalous acceleration experienced by spacecrafts in the solar system,
all suggest a breakdown of Newton's law of gravity in the weak 
acceleration regime.

\begin{figure}
\plotone{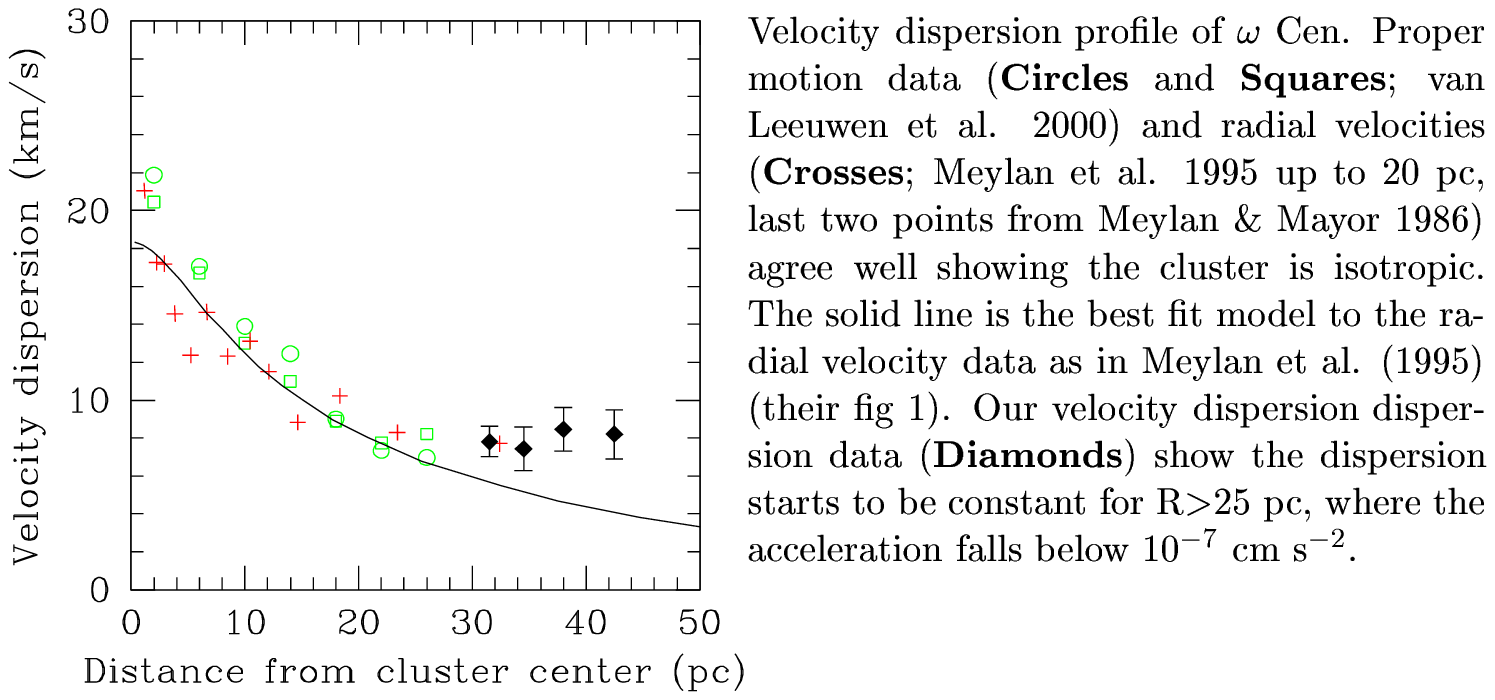}
\end{figure}

\end{document}